\documentclass[review]{elsarticle}
\usepackage{tabularx}
\usepackage[utf8]{inputenc}
\usepackage{lineno,hyperref}
\usepackage[top=1.65in, bottom=1.65in, left=1.35in, right=1.35in]{geometry}
\usepackage{setspace}
\usepackage{amsmath}
\usepackage{tikz}
\usetikzlibrary{positioning, arrows, fit}
\usepackage{graphicx}
\usepackage{threeparttable}
\usepackage{float}

\biboptions{numbers}

\makeatletter
\def\ps@pprintTitle{%
 \let\@oddhead\@empty
 \let\@evenhead\@empty
 \def\@oddfoot{}%
 \let\@evenfoot\@oddfoot}
\makeatother
\begin{document}

\begin{frontmatter}
\title{Evaluating Small-Scale Code Models for Code Clone Detection}
\author{Jorge Martinez-Gil}
\address{Software Competence Center Hagenberg GmbH \\ Softwarepark 32a, 4232 Hagenberg, Austria \\ \url{jorge.martinez-gil@scch.at}}

\begin{abstract}
Detecting code clones is relevant to software maintenance and code refactoring. This challenge still presents unresolved cases, mainly when structural similarity does not reflect functional equivalence, though recent code models show promise. Therefore, this research aims to systematically measure the performance of several newly introduced small code models in classifying code pairs as clones or non-clones. The evaluation is based on five datasets: BigCloneBench, CodeJam, Karnalim, POJ104, and PoolC, as well as six code models: CodeBERT, GraphCodeBERT, Salesforce T5, UniXCoder, PLBART, and Polycoder. Most models performed well across standard metrics, including accuracy, precision, recall, and F1-score. However, a marginal fraction of clones remains challenging to detect, especially when the code looks similar but performs different operations. The source code that illustrates our approach is available at:
\url{https://github.com/jorge-martinez-gil/small-code-models}
\end{abstract}


\end{frontmatter}

\section{Introduction}
The presence of code clones is a frequent issue in software development, affecting maintainability. Code duplication can raise complexity, increase maintenance costs, and introduce mistakes during updates. Therefore, clone detection helps developers find and restructure redundant code snippets, which usually supports better software organization \cite{rattan2013software}.

Earlier approaches to clone detection are usually grouped into three types: token-based, tree-based, and metric-based. Token-based methods compare sequences of tokens to detect similar codes. Tree-based techniques rely on abstract syntax trees to assess structural likeness, and metric-based ones convert code into numerical features for similarity checks. These techniques were generally effective at finding exact copies or structurally related code but often miss clones that perform the same task with different syntax.

Over the last ten years, machine learning (ML) has brought new approaches to clone detection. For example, some deep learning (DL) models trained on extensive code collections have shown the ability to recognize patterns that go beyond surface-level similarity \cite{karmakar2021pre}. These models learn behavioral aspects of code, allowing them to identify matches that traditional techniques often overlook. Their strong performance led them to dominate benchmark rankings soon after their introduction.

Recently, new code models have been introduced that demonstrate improved performance \cite{almatrafi2025code}. Given the increasing awareness of large models' environmental and computational costs, it is worth investigating whether smaller, more resource-efficient models can achieve results comparable to the larger ones. For this reason, this study focuses on models considered small, acknowledging that there is no universally accepted definition of a small code model. In line with current developments, we define small-scale models as those at least two orders of magnitude smaller than the largest available code models (medium models would be roughly one order smaller). The number of parameters is typically used as the basis for such comparisons. Therefore, we consider models with just a few hundred million parameters in practice. These smaller models present a trade-off between performance and efficiency, making them suitable for use in environments with limited computational resources.

So far, there has been no thorough comparative and exhaustive evaluation of how well small code models perform in clone detection. Therefore, this work is the first to compare multiple small-scale code models for clone detection across several benchmark datasets using a consistent evaluation setup. The goal is to answer the following research questions:

\begin{itemize}
	\item \textbf{RQ1:} What is the performance of small code models in clone detection on the most adopted clone-related datasets?
	\item \textbf{RQ2:} What small code model demonstrates the highest reliability?
	\item \textbf{RQ3:} Which architectural features in small models contribute most to clone detection performance?
\end{itemize}

The rest of this paper is structured as follows: Section 2 reviews related work, covering traditional and modern approaches for clone detection. Section 3 presents the methodology, including dataset details, model architecture, and evaluation metrics that we have used. Section 4 shows our experimental results, analyzing the performance of several models over different datasets under equivalent conditions. Section 5 discusses the findings and potential limitations. Section 6 concludes with a summary of contributions and possible directions for future research.

\section{Related Work}
The problem of the code clones or equivalent code snippets is common in software development \cite{zakeri2023systematic}. They can complicate maintenance, as changes or bug fixes in one instance often need to be replicated in others. Clones may also make it difficult to understand the programs and contribute to inconsistencies or defect propagation across large codebases \cite{martinez2024evaluation}.

Code clones are usually grouped into four classes: Type-1 (identical code), Type-2 (minor edits like renaming), Type-3 (structural changes with retained logic), and Type-4 (different structure but same functionality). Traditional methods perform well on Type-1, Type-2, and Type-3, but Type-4 remains challenging due to limited surface similarity. This limitation has increased interest in code models capable of identifying deeper similarities \cite{key-KaurR23}. Smaller models that encode structure and meaning have shown promising results in identifying such clones, mainly when computing resources are restricted. Their ability to work across different programming languages makes them practical for automated code maintenance \cite{zhang2023language}.

\subsection{Code-related tasks, clone detection, and their applications}
Recent work has extended the use of language models to code-related tasks beyond generation. Pham et al.\cite{DBLP:journals/access/PhamHTHTL24} introduced MAGECODE, a system for detecting machine-generated code using pre-trained models that can distinguish between human- and machine-written code. Hemberg et al.\cite{DBLP:journals/gpem/HembergMO24} applied language models to program refinement, showing their use in automated code modification. Husein et al.\cite{DBLP:journals/csi/HuseinAC25} reviewed the performance of language models in code completion, cataloging their current capabilities across languages and tasks. Ramler et al.\cite{DBLP:conf/llm4code/RamlerM0NH24} documented how AI-assisted coding tools are used in professional software development. Finally, Zhang et al.\cite{DBLP:journals/ijseke/ZhangCCCZ24} explored multi-intent code comment generation using large models, aiming to support code comprehension and maintainability in day-to-day programming. 

Clone detection has a long research tradition as a subfield of code-related tasks. Proposed methods are classified as lexical, syntactic, or semantic, with the semantic category being the most difficult to model. Prior work \cite{key-martinez-swqd} has investigated unsupervised techniques based on multiple similarity metrics. Traditional methods are effective for exact or near-exact matches, whereas neural models offer better recall by identifying functionally similar code despite structural variation. Early neural approaches like code2vec \cite{alon2019code2vec} showed that embeddings can capture structure and behavior. More recent efforts explore hybrid models that use genetic programming \cite{key-martinez-ijseke}, and ensemble-based unsupervised approaches \cite{key-martinez-codesim} continue to show potential for integration with learned models. Semantic similarity remains challenging to quantify objectively \cite{haque2022semantic} and interpretability as well \cite{martinez2024augmenting}. The reason is the wide variety of applications where it can have an impact. 

\subsection{Code Models}
Transformer-based models such as BERT \cite{key-Bert} introduced pretraining strategies adapted to source code, contributing to improvements in code-related tasks. Recent work \cite{zhang2025machine} has analyzed the internal representations learned from code, showing that transformer architectures are particularly effective. Table~\ref{tab:clone_models} lists six models commonly used in this area.

\begin{table}[h]
    \centering
    \begin{threeparttable}
    \begin{tabular}{|l|c|c|p{5cm}|}
        \hline
        \textbf{Model} & \textbf{Architecture} & \textbf{Parameters} & \textbf{Significance} \\
        \hline
        CodeBERT \cite{key-codebert} & BERT-based  & 125M & Transformer-based model pre-trained on source code and natural language, effective for code similarity and clone detection. \\
        \hline
        GraphCodeBERT \cite{guo2020graphcodebert} & BERT+graphs & 125M & Extension of CodeBERT by incorporating data flow information, improving structural understanding in code-related tasks. \\
        \hline
        Salesforce T5 (base) \cite{wang2021codet5} & Seq2Seq  & 220M  & A T5-based model fine-tuned for code-related NLP tasks, including clone detection and code summarization. \\
        \hline
        UniXCoder \cite{guo2022unixcoder} & Encoder-Decoder & undisclosed\textsuperscript{*} & A multi-modal model that uses both token and structure embeddings, enhancing performance on various code intelligence tasks. \\
        \hline
        PLBART \cite{ahmad2021unified} & Seq2Seq & 140M & Pre-trained model suitable for generation and classification tasks, including clone detection. \\
        \hline
        PolyCoder (base) \cite{xu2022systematic} & Decoder-only  & 160M & Model designed for code generation and understanding, trained on multiple programming languages, making it versatile for clone detection. \\
        \hline
    \end{tabular}
    \begin{tablenotes}
        \item[*] Although the authors have not disclosed the number of parameters, its physical size suggests that it is below 200M.
    \end{tablenotes}
    \caption{Clone Detection Models, Architectures, Parameters, and Their Significance}
    \label{tab:clone_models}
    \end{threeparttable}
\end{table}

These models process source code as token sequences and consider structural and semantic aspects. Current efforts aim to make these models more effective when dealing with complex code structures.

\subsection{Powerful Large Code Models vs Efficient Small Code Models}
Small models are often better when hardware limits are strict or low latency is essential. Typical cases include code editors, browser extensions, mobile development environments, or continuous integration pipelines. These systems require quick responses and minimal resource usage. A smaller model can typically run without a GPU and still return valuable results, making integrating it into everyday developer tools easier.

They can also work as filters in multi-stage setups. The smaller model handles the easy cases and forwards only uncertain or borderline inputs to a larger model. This setup can reduce total computation time without sacrificing too much accuracy. It also avoids the energy demands of running large models constantly, which matters in shared environments.

\subsection{Contribution Over the State-of-the-Art}
This work compares several small code models used for clone detection. Many models have been released recently, but direct comparisons under the same setup are rare. This study addresses that by evaluating a diverse group of architectures under consistent conditions. CodeBERT and GraphCodeBERT are encoder-only and are mainly used for classification tasks. CodeT5 and PLBART use a sequence-to-sequence structure suitable for generation and classification. UniXCoder supports both task types without needing changes to its architecture. PolyCoder is decoder-only and belongs to the GPT family.

These models differ in how they represent source code. CodeBERT uses plain text tokens. GraphCodeBERT and UniXCoder include structural elements, which help when surface similarity is insufficient. PLBART and PolyCoder are geared toward code generation, with PolyCoder also trained in multiple programming languages. Among the models considered, Salesforce T5 has the highest parameter count.

Please note that in the context of this study, we have explicitly discarded CodeParrot \cite{tunstall2022natural} and CuBERT \cite{kanade2020learning} due to their focus on Python-only code, which limits generalization across multiple programming languages. Additionally, we have excluded some very popular models such as StarCoder \cite{li2023starcoder}, Starcoder2 \cite{lozhkov2024starcoder}, Incoder \cite{fried2022incoder}, and CodeLlama \cite{roziere2023code} because their most basic versions range between 3B and 7B parameters, which cannot guarantee a balanced comparison with the models with significantly lower parameter counts we are considering here.

\section{Methodology}
There has been limited evaluation of small code models in clone detection so far \cite{ZhaoLT0YL025}. To address this gap, we ran experiments using five well-known datasets: BigCloneBench \cite{svajlenko2021bigclonebench}, CodeJam \cite{zhao2018deepsim}, Karnalim \cite{karnalim2019source}, PoolC \cite{mou2016convolutional}, and POJ104 \cite{poolc}. Each dataset includes (implicitly or explicitly) labeled code pairs, making measuring model performance across different types of clones possible. The six models selected for this study were CodeBERT, GraphCodeBERT, Salesforce T5, UniXCoder, PLBART, and PolyCoder. They were chosen based on their parameter count and availability. All models were tested under the same setup using accuracy, precision, recall, and F1-score.

\subsection{Technical Features}
Table~\ref{tab:clone_datasets} lists the datasets used to measure clone detection performance, including sample counts and the programming languages they cover. The datasets differ in source, size, and purpose, allowing for evaluation across various conditions.

\begin{table}[h]
    \centering
    \begin{tabular}{|l|c|c|c|}
        \hline
        \textbf{Dataset} & \textbf{Training Samples} & \textbf{Test Samples} & \textbf{Languages} \\
        \hline
        BigCloneBench  & 900k & 415k & Java \\
        CodeJam  & 726k & 14k & Java \\
        Karnalim  & 327 & 140 & Java \\
        POJ104  & 32k & 12k & C++ \\
        PoolC  & 480k & 120k & Python \\
        \hline
    \end{tabular}
    \caption{Clone Detection Datasets: Samples and Programming Languages}
    \label{tab:clone_datasets}
\end{table}

\subsection{The BigCloneBench Dataset} 
BigCloneBench (BCB) is a large dataset built from open-source projects and labeled with clone types ranging from exact copies to structurally different but functionally similar code. Labels are generated using a mix of automated tools and manual checks. It is widely used to test clone detection methods, including older techniques and ML models. Due to its size, it is also suitable for testing how well models scale.

\subsection{The Google Code Jam Dataset}
The Google Code Jam (GCJ) dataset includes code submissions from the Google Code Jam contest, in which many participants solved the same problems. This led to multiple implementations of similar algorithms, differing in structure, style, and efficiency. Such variation allows models to be evaluated on functional similarity rather than exact matches, making the task harder and more realistic.

\subsection{The Karnalim Dataset}
The Karnalim dataset is used for source code plagiarism detection in academic settings. It contains labeled pairs of student submissions, some modified copies of others. Many samples include changes such as renaming variables or altering code structure to avoid easy detection. This makes it suitable for testing models that need to go beyond simple pattern matching. While smaller than datasets like BigCloneBench, it is well-suited for evaluating techniques to detect disguised reuse in student work.

\subsection{The Peking University Online Judge Dataset}
The POJ104 dataset includes student code submissions from the Peking University Online Judge. Each sample is tied to a specific programming problem, which allows grouping based on intended functionality. The dataset covers various styles and coding strategies, making it useful for testing models that aim to detect functional similarity. While it does not contain direct clone labels, the structure supports indirectly evaluating clone detection methods.

\subsection{Poolc Benchmark Dataset}  
The PoolC dataset contains labeled code pairs marked as clones or non-clones. It includes samples from open-source projects covering a range of programming styles. Clone types include near duplicates, minor edits, and structurally different but functionally similar code. PoolC is split into training, validation, and test sets, making it suitable for ML experiments. Its structure allows consistent comparisons across models, though varying styles and unclear boundaries between clone types add difficulty.

\section{Empirical Evaluation}
This section presents an empirical evaluation of the small code models, focusing on their training setup, evaluation metrics, and experimental results. The goal is to understand the model's strengths and limitations across different datasets and identify areas for improvement.

\subsection{Model and Training Setup}
All the models have been fine-tuned for binary classification (clones or non-clones). The idea is that the models process concatenated code snippets and use attention mechanisms to determine similarity. The training parameters that we have used in all cases are:

\begin{itemize}
    \item Batch size: 8
    \item Epochs: 3
    \item Weight decay: 0.01
    \item Seed: 42
		\item Optimization for: F1-score
\end{itemize}

Please note that we allow each model's learning rate to be the one by default.

\subsection{Evaluation Metrics}
Performance is measured using four metrics: accuracy, which refers to the proportion of correctly classified code snippets; precision, which indicates the proportion of detected clones that are actual clones; recall, which captures the proportion of actual clones that are correctly detected; and F1-score, defined as the harmonic mean of precision and recall.

\subsection{BCB Dataset} 
Table \ref{tab:bcb} presents the performance evaluation of our small-scale code models on the BCB benchmark dataset.

\begin{table}[H]
    \centering
    \begin{tabular}{|l|c|c|c|c|}
        \hline
        \textbf{Benchmark} & \textbf{Accuracy} & \textbf{Precision} & \textbf{Recall} & \textbf{F1-score} \\
        \hline
        CodeBERT  		& 0.954	& 0.798	& 0.897	& 0.844      \\
        GraphCodeBERT & 0.959	& 0.832	& 0.888	& 0.858      \\
        Salesforce T5	& 0.970	& 0.868	& 0.926	& 0.896      \\
        UniXCoder   	& 0.971	&	0.864	& 0.941	& 0.901      \\
				PLBART   			& 0.973	&	0.876	& 0.936	& 0.905       \\
				PolyCoder   	& 0.951	&	0.798	& 0.874	& 0.834      \\
        \hline
    \end{tabular}
    \caption{Performance on the BCB benchmark dataset}
    \label{tab:bcb}
\end{table}

\begin{itemize}
    \item PLBART achieves the best performance, with the highest accuracy (0.973), precision (0.876), and a strong F1-score (0.905). Therefore, PLBART is the most balanced model for detecting code clones in BCB.
    \item UniXCoder also performs exceptionally well, achieving an accuracy of 0.971 and the highest recall (0.941). Therefore, UniXCoder is highly effective at identifying true clones but has slightly lower precision (0.864) than PLBART.
    \item Salesforce T5 achieves a strong F1-score of 0.896, with good precision (0.868) and recall (0.926). This model performs well but does not surpass PLBART or UniXCoder in performance.
    \item GraphCodeBERT and CodeBERT perform moderately, with GraphCodeBERT showing better precision (0.832) than CodeBERT (0.798). However, their recall and F1 scores are significantly lower than the top-performing models.
    \item PolyCoder has the weakest performance, with the lowest accuracy (0.951) and F1-score (0.834). 
\end{itemize}

\subsection{GCJ Dataset}
Table \ref{tab:gcj} shows the performance of the small-scale code models under study on the GCJ benchmark dataset.

\begin{table}[h]
    \centering
    \begin{tabular}{|l|c|c|c|c|}
        \hline
        \textbf{Benchmark} & \textbf{Accuracy} & \textbf{Precision} & \textbf{Recall} & \textbf{F1-score} \\
        \hline
        CodeBERT  		& 1.000	& 1.000	& 1.000	& 1.000      \\
        GraphCodeBERT & 1.000	& 1.000	& 1.000	& 1.000      \\
        Salesforce T5	&	1.000	& 1.000	& 1.000	& 1.000      \\
        UniXCoder   	& 1.000	& 1.000	& 1.000	& 1.000      \\
				PLBART   			& 1.000	& 1.000	& 1.000	& 1.000       \\
				PolyCoder  		& 1.000	& 1.000	& 1.000	& 1.000       \\
        \hline
    \end{tabular}
    \caption{Performance on the GCJ benchmark dataset}
    \label{tab:gcj}
\end{table}

\begin{itemize} 
	\item All models achieve perfect performance on the GCJ dataset, with accuracy, precision, recall, and F1-score equal to 1.000.
	\item Such good results may indicate a problem (data bias or limited sample variation), but after further investigation, we have found that such results can be achieved with sufficient training. When training with less data, these results are not achieved.
	\item These results may reflect unknown dataset-specific characteristics, making it easier for these models to generalize. 
\end{itemize}

\subsection{Karnalim Dataset}
Table \ref{tab:karnalim} reports the performance of our small-scale code models on the Karnalim benchmark dataset.

\begin{table}[h]
    \centering
    \begin{tabular}{|l|c|c|c|c|}
        \hline
        \textbf{Benchmark} & \textbf{Accuracy} & \textbf{Precision} & \textbf{Recall} & \textbf{F1-score} \\
        \hline
        CodeBERT  		& 0.667	& 0.667	& 1.000	& 0.800      \\
        GraphCodeBERT & 0.899	& 0.882	& 0.978	& 0.923      \\
        Salesforce T5	&	0.725	& 0.708	& 1.000	& 0.829      \\
        UniXCoder   	& 0.942	&	0.938	& 0.978	& 0.957      \\
				PLBART   			& 0.971	& 0.978	& 0.978	& 0.978      \\
				PolyCoder  		& 0.855	&	0.833	& 0.978	& 0.900      \\
        \hline
    \end{tabular}
    \caption{Performance on the Karnalim benchmark dataset}
    \label{tab:karnalim}
\end{table}

\begin{itemize} 
	\item PLBART achieves the best performance, with the highest accuracy (0.971) and F1-score (0.978). 
	\item UniXCoder also performs strongly, reaching 0.942 accuracy and 0.957 F1-score, with a slight drop in precision compared to PLBART. 
	\item GraphCodeBERT follows, with solid scores across all metrics, particularly an F1-score of 0.923 and an accuracy of 0.899. 
	\item PolyCoder performs reasonably well, with high recall (0.978) but a lower precision (0.833). 
	\item Salesforce T5 and CodeBERT lag behind, achieving perfect recall (1.000) but low precision (0.708 and 0.667), leading to a noticeable drop in accuracy and F1-score. 
\end{itemize}

\subsection{POJ104 Dataset}
Table \ref{tab:poj104} reports how our small-scale code models under study can perform on the POJ104 dataset.

\begin{table}[h]
    \centering
    \begin{tabular}{|l|c|c|c|c|}
        \hline
        \textbf{Benchmark} & \textbf{Accuracy} & \textbf{Precision} & \textbf{Recall} & \textbf{F1-score} \\
        \hline
        CodeBERT  		& 0.863	& 0.856	& 0.867	& 0.861      \\
        GraphCodeBERT & 0.864	& 0.857	& 0.867	& 0.862      \\
        Salesforce T5 & 0.600	& 0.567	& 0.794	& 0.662      \\
        UniXCoder   	& 0.883	&	0.883	& 0.879	& 0.881      \\
				PLBART   			& 0.733	&	0.676	& 0.875	& 0.763      \\
				PolyCoder   	& 0.900	&	0.895	& 0.905	& 0.900      \\
        \hline
    \end{tabular}
    \caption{Performance on the POJ104 benchmark dataset}
    \label{tab:poj104}
\end{table}

\begin{itemize} 
	\item PolyCoder achieves the best performance, with accuracy and F1-score of 0.900, as well as balanced precision (0.895) and recall (0.905). 
	\item UniXCoder also performs strongly, reaching 0.883 accuracy and 0.881 F1-score, maintaining a close balance between precision and recall. 
	\item GraphCodeBERT and CodeBERT show solid and consistent results, with nearly identical metrics and F1-scores of 0.862 and 0.861, respectively. 
	\item PLBART shows acceptable recall (0.875) but has noticeably lower precision (0.676), resulting in a moderate F1-score of 0.763. 
	\item Salesforce T5 ranks lowest, with the weakest accuracy (0.600) and lowest precision (0.567), indicating a high rate of false positives despite a decent recall (0.794). 
\end{itemize}

\subsection{PoolC Dataset} 
Table \ref{tab:poolc} presents the performance evaluation of our small-scale code models on the PoolC benchmark dataset. 

\begin{table}[h]
    \centering
    \begin{tabular}{|l|c|c|c|c|}
        \hline
        \textbf{Benchmark} & \textbf{Accuracy} & \textbf{Precision} & \textbf{Recall} & \textbf{F1-score} \\
        \hline
        CodeBERT  		& 0.936	& 0.934	& 0.938	& 0.936      \\
        GraphCodeBERT & 0.942	& 0.934	& 0.949	& 0.941      \\
        Salesforce T5 & 0.943	& 0.914	& 0.977	& 0.944      \\
        UniXCoder   	& 0.949	&	0.960	& 0.936	& 0.948      \\
				PLBART   			& 0.937	&	0.929	& 0.946	& 0.937      \\
				PolyCoder   	& 0.924	&	0.912	& 0.936	& 0.924      \\
        \hline
    \end{tabular}
    \caption{Performance on the PoolC benchmark dataset}
    \label{tab:poolc}
\end{table}

\begin{itemize} 
	\item UniXCoder achieves the best performance, with the highest F1-score (0.948), precision (0.960), and strong accuracy (0.949). 
	\item Salesforce T5 shows the highest recall (0.977) and strong F1-score (0.944), though its lower precision (0.914) indicates a higher rate of false positives. 
	\item GraphCodeBERT also performs well, with a good balance of recall (0.949), precision (0.934), and F1-score (0.941). 
	\item CodeBERT and PLBART are slightly behind, reaching F1-scores of 0.936–0.937, with minor differences in recall and precision. 
	\item PolyCoder scores lowest, with accuracy and F1-score of 0.924 due to its lower precision (0.912). 
\end{itemize}

\subsection{Summary}
\begin{figure}[ht]
    \centering
    \includegraphics[width=0.7\textwidth]{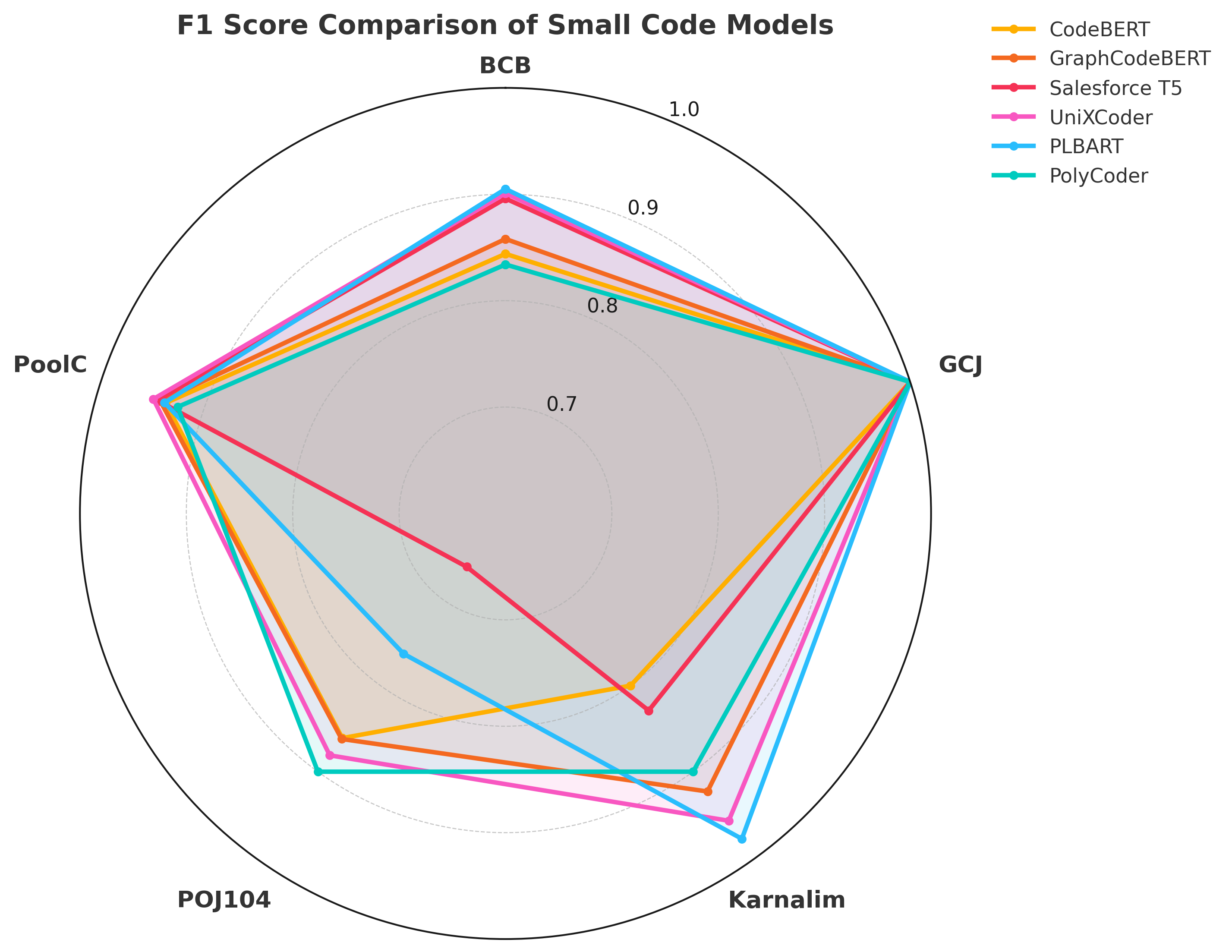}
    \caption{F1 scores of small code models measured on five standard clone detection datasets. Each axis represents one dataset, and each line tracks the performance of one model across all datasets.}
    \label{fig:f1_score_models}
\end{figure}

Figure~\ref{fig:f1_score_models} shows the performance distribution of the six small-scale code models across the five benchmarks. The plot allows a quick comparison of their relative strengths. UniXCoder maintains a stable performance profile and reaches high scores across the board. PLBART performs well on most datasets, though it drops on POJ104. Salesforce T5 shows higher variance, with a lower score on POJ104 too. CodeBERT and PolyCoder achieve reasonable results but are slightly behind in consistency.

Table~\ref{tab:rankings} shows the F1-score rankings of six code models across five benchmark datasets. A lower rank indicates better performance, with one being the best. Each entry reflects the position of a model on a specific dataset based on its F1 score.  PLBART ranks in three datasets and stays within the top five. UniXCoder also performs well, never ranking lower than second place across all datasets, while other models show greater variation in their positions.

\begin{table}[h]
\centering
\begin{tabular}{|l|c|c|c|c|c|}
\hline
\textbf{Model} & \textbf{BCB} & \textbf{GCJ} & \textbf{Karnalim} & \textbf{POJ104} & \textbf{PoolC} \\
\hline
CodeBERT        & 5 & \textbf{1} & 6 & 4 & 4 \\
GraphCodeBERT   & 4 & \textbf{1} & 3 & 3 & 3 \\
Salesforce T5   & 3 & \textbf{1} & 5 & 6 & 2 \\
UniXCoder       & 2 & \textbf{1} & 2 & 2 & \textbf{1} \\
PLBART          & \textbf{1} & \textbf{1} & \textbf{1} & 5 & 4 \\
PolyCoder       & 6 & \textbf{1} & 4 & \textbf{1} & 6 \\
\hline
\end{tabular}
\caption{F1-score ranking (1 = best) of each model across all benchmark datasets}
\label{tab:rankings}
\end{table}

Table~\ref{tab:summary} shows the mean and standard deviation of the F1-score across datasets. UniXCoder performs well and is the most stable. GraphCodeBERT has the highest mean but with slightly more fluctuation. PLBART and Salesforce T5 vary more across datasets, while PolyCoder is more uniform but has lower averages.

\begin{table}[h]
\centering
\begin{tabular}{|l|c|c|}
\hline
\textbf{Model} & \textbf{Mean F1-score} & \textbf{Standard Deviation} \\
\hline
GraphCodeBERT   & 0.925 & 0.053 \\
UniXCoder       & 0.918 & 0.041 \\
PLBART          & 0.904 & 0.083 \\
CodeBERT        & 0.895 & 0.072 \\
PolyCoder       & 0.892 & 0.030 \\
Salesforce T5   & 0.855 & 0.137 \\
\hline
\end{tabular}
\caption{Mean and standard deviation of F1-score across all benchmark datasets}
\label{tab:summary}
\end{table}

Table~\ref{tab:summary_equal} shows each model's F1-score mean and standard deviation, assuming the validation instances solved from each benchmark dataset. This is done to decrease the importance of small datasets such as Karnalim when establishing the ranking. UniXCoder leads with the highest average (0.937) and the lowest standard deviation (0.042). GraphCodeBERT and PLBART follow, averaging 0.917, though PLBART shows more variation. PolyCoder performs slightly below, while CodeBERT and Salesforce T5 rank lower.

\begin{table}[h]
\centering
\begin{tabular}{|l|c|c|}
\hline
\textbf{Model} & \textbf{E.-W. Mean F1-score} & \textbf{E.-W. Std. Dev.} \\
\hline
UniXCoder       & 0.937 & 0.042 \\
GraphCodeBERT   & 0.917 & 0.053 \\
PLBART          & 0.917 & 0.084 \\
PolyCoder       & 0.912 & 0.053 \\
CodeBERT        & 0.888 & 0.071 \\
Salesforce T5   & 0.866 & 0.126 \\
\hline
\end{tabular}
\caption{Mean and standard deviation of F1-score across all benchmark datasets (equal-weighted, instance count)}
\label{tab:summary_equal}
\end{table}

\subsection{Answer to the Research Questions}
As a conclusion of our experiments, we can now answer the research questions previously formulated. \\ \\
\noindent \textbf{RQ1: What is the performance of small code models in clone detection on the most adopted clone-related datasets?} \\
\noindent \textbf{AQ1:} GraphCodeBERT reaches the highest average F1-score (0.925) and ranks consistently high in most datasets. PLBART and UniXCoder follow closely. However, if we consider the total number of instances validated, UniXCoder reaches the highest average F1-score (0.937). GraphCodeBERT and PLBART follow closely, both averaging 0.917.\\

\noindent \textbf{RQ2: What small code model demonstrates the highest reliability?} \\
\noindent \textbf{AQ2:} UniXCoder has a high mean F1-score (0.918) and the lowest standard deviation (0.041), which means stable performance across benchmarks. This is also valid even if all the instances validated are counted equally.\\

\noindent \textbf{RQ3: Which architectural features in small models contribute most to clone detection performance?} \\
\noindent \textbf{AQ3:} Graph-based representations (GraphCodeBERT) and multi-modal encoding (UniXCoder) perform better when compared to models that rely on general-purpose sequence-to-sequence designs. Models like Salesforce T5 and PLBART show variable results across datasets, and CodeBERT and PolyCoder generally score lower. This means specialized code-aware pretraining benefits clone detection more than general sequence modeling. \\

Furthermore, Table \ref{tab:model-recommendations} summarizes the practical strengths of each small code model and suggests usage scenarios where their design or performance profile is particularly well-suited.

\begin{table}[ht]
\centering
\begin{tabular}{|p{4cm}|p{3.4cm}|p{6cm}|}
\hline
\textbf{Use Case} & \textbf{Recommended Model(s)} & \textbf{Rationale} \\
\hline
High recall (catch most clones) & PLBART, Salesforce T5 & Near-perfect recall on multiple datasets; suitable when missing a clone is costly. \\
\hline
Low false positives (high precision) & UniXCoder, GraphCodeBERT & Strong precision and F1 scores; ideal when clone over-identification must be avoided. \\
\hline
Balanced performance (general use) & UniXCoder, PLBART & High F1 scores across datasets with good precision-recall trade-offs. \\
\hline
Resource-constrained environments (no GPU) & CodeBERT, PolyCoder & Fewer parameters; acceptable results without high compute requirements. \\
\hline
Cross-language or mixed-language projects & UniXCoder, PolyCoder & Multi-language support; performs well across language-diverse datasets. \\
\hline
Educational plagiarism detection & GraphCodeBERT, PLBART & Ideal when structural changes are present; high scores on the Karnalim dataset. \\
\hline
Real-time tools (low latency prediction) & CodeBERT & Encoder-only model; lightweight with fast inference time. \\
\hline
Multi-stage pipeline (filter + rerank) & CodeBERT (filter), PLBART (rerank) & Efficient two-stage setup: CodeBERT filters obvious cases, PLBART handles borderline examples. \\
\hline
\end{tabular}
\caption{Recommended small-scale code models by use case}
\label{tab:model-recommendations}
\end{table}

\section{Discussion}
Results allow us to compare model behavior across datasets and identify where models tend to perform well and where they do not. Each dataset tests different qualities, making it possible to observe differences in accuracy or consistency. Some models perform well across most cases, while others are more sensitive to specific dataset characteristics. 

\subsection {Observed Strengths}
Our observations reveal that the models we have considered can maintain high accuracy across different datasets. Furthermore, the models generalize well to unseen examples. While most models perform well in most cases, there are still some problems with specific datasets that exhibit different types of clones or have less training data. 

The evaluation across multiple datasets reveals that UniXCoder and GraphCodeBERT achieve strong performance, often ranking among the top models. UniXCoder leads on the PoolC and POJ104 datasets and remains highly competitive on BCB and Karnalim. It combines high recall with solid precision, effectively controlling false positives. GraphCodeBERT also shows strong generalization capabilities, particularly excelling in Karnalim. 

In contrast, the performance of PLBART, Salesforce T5, and PolyCoder varies across datasets. While PLBART shines on BCB with top F1 scores, it underperforms in POJ104 and GCJ. Salesforce T5, although achieving top scores on GCJ, shows inconsistent behavior, especially in POJ104 and Karnalim, where its accuracy drops sharply. PolyCoder performs moderately, with lower precision and recall. These inconsistencies show the importance of evaluating models on diverse datasets, as performance can vary widely depending on data characteristics.

\subsection{Observed Limitations}
The models tested in this study show strong performance on several datasets, but there are still areas where they fall short. In particular, datasets with code snippets that appear similar in structure but differ in behavior often lead to wrong predictions. This means that the models favor surface patterns over deeper meaning. 

Additionally, some models, including PLBART and Salesforce T5, show high recall but lower precision on specific datasets, indicating a tendency to classify too many pairs as clones. Small changes in the syntax, like alternative loop styles, can still mislead some models, particularly those that do not use structural information.

\subsection{Threats to Validity}
Our results must be interpreted carefully, as several factors may limit the validity of direct comparisons between models. While efforts were made to maintain consistency, some decisions, like using fixed hyperparameters or fixed random seeds, may favor some models over the rest. The following points outline potential limitations that could affect the fairness of the reported results.

\begin{itemize}
    \item Several small-scale code models included in the benchmark were not originally trained for classification tasks. As a result, their internal representations and optimization objectives might not be optimal for clone detection, which may affect their performance.
    
    \item All models were trained using the same set of hyperparameters. While this favors consistency, it does not account for the possibility that different models may require different configurations to achieve optimal performance. This uniform setup could disadvantage some models and lead to unfair comparisons.
    
    \item We do not know if any small code model has been pre-trained with test data from any benchmarks considered. If so, the comparison would not be fair either.
		
		 \item The same random seed was used across all training runs to facilitate reproducibility. However, this may also constrain the variability in training outcomes, which could be particularly relevant for models with high sensitivity to initialization.

    \item The optimization objective was set to maximize the F1 score. While this is a common choice, it implicitly balances precision and recall equally, which may not align with the priorities of all clone detection use cases. 
\end{itemize}

\section{Conclusions}
This research has evaluated a selection of small pre-trained code models for automated clone detection across diverse benchmark datasets. When trained appropriately, the results demonstrate that transformer-based models can deliver high clone detection performance, even under resource-constrained conditions. For example, models such as GraphCodeBERT and UniXCoder have consistently achieved high accuracy and F1 scores across datasets, confirming that compact models can be both practical and efficient in identifying code similarity.

The analysis also revealed some challenges in detecting semantically equivalent clones that differ in syntactic structure. While most models performed well in identifying syntactically similar clones, their performance degraded in datasets involving deeper logical variations, pointing out some limitations of representation learning strategies. These findings indicate the need for novel ways to better capture code intent, functionality, and control flow beyond surface-level patterns.

Future work must explore novel ways to measure how performance changes with code model size and training time, which is relevant for deployment in environments with limited hardware. Moreover, testing with adversarial code will also help assess performance under less controlled conditions.

\section*{Acknowledgments}
The research reported in this paper has been funded by the Federal Ministry for Climate Action, Environment, Energy, Mobility, Innovation, and Technology (BMK), the Federal Ministry for Digital and Economic Affairs (BMDW), and the State of Upper Austria in the frame of SCCH, a center in the COMET - Competence Centers for Excellent Technologies Programme.

\bibliography{mybib}

\begin{thebibliography}{38}
\expandafter\ifx\csname natexlab\endcsname\relax\def\natexlab#1{#1}\fi
\providecommand{\url}[1]{\texttt{#1}}
\providecommand{\href}[2]{#2}
\providecommand{\path}[1]{#1}
\providecommand{\DOIprefix}{doi:}
\providecommand{\ArXivprefix}{arXiv:}
\providecommand{\URLprefix}{URL: }
\providecommand{\Pubmedprefix}{pmid:}
\providecommand{\doi}[1]{\href{http://dx.doi.org/#1}{\path{#1}}}
\providecommand{\Pubmed}[1]{\href{pmid:#1}{\path{#1}}}
\providecommand{\bibinfo}[2]{#2}
\ifx\xfnm\relax \def\xfnm[#1]{\unskip,\space#1}\fi
\bibitem[{Ahmad et~al.(2021)Ahmad, Chakraborty, Ray \&
  Chang}]{ahmad2021unified}
\bibinfo{author}{Ahmad, W.~U.}, \bibinfo{author}{Chakraborty, S.},
  \bibinfo{author}{Ray, B.}, \& \bibinfo{author}{Chang, K.}
  (\bibinfo{year}{2021}).
\newblock \bibinfo{title}{Unified pre-training for program understanding and
  generation}, .
\newblock (pp. \bibinfo{pages}{2655--2668}).
  \DOIprefix\doi{10.18653/V1/2021.NAACL-MAIN.211}.
\bibitem[{Almatrafi et~al.(2025)Almatrafi, Eassa \& Sharaf}]{almatrafi2025code}
\bibinfo{author}{Almatrafi, A.~A.}, \bibinfo{author}{Eassa, F.~A.}, \&
  \bibinfo{author}{Sharaf, S.~A.} (\bibinfo{year}{2025}).
\newblock \bibinfo{title}{Code clone detection techniques based on large
  language models}.
\newblock {\it \bibinfo{journal}{IEEE Access}\/}, .
  \DOIprefix\doi{10.1109/ACCESS.2025.3549780}.
\bibitem[{Alon et~al.(2019)Alon, Zilberstein, Levy \& Yahav}]{alon2019code2vec}
\bibinfo{author}{Alon, U.}, \bibinfo{author}{Zilberstein, M.},
  \bibinfo{author}{Levy, O.}, \& \bibinfo{author}{Yahav, E.}
  (\bibinfo{year}{2019}).
\newblock \bibinfo{title}{code2vec: Learning distributed representations of
  code}.
\newblock {\it \bibinfo{journal}{Proceedings of the ACM on Programming
  Languages}\/},  {\it \bibinfo{volume}{3}\/}, \bibinfo{pages}{1--29}.
  \DOIprefix\doi{10.1145/3290353}.
\bibitem[{Devlin et~al.(2019)Devlin, Chang, Lee \& Toutanova}]{key-Bert}
\bibinfo{author}{Devlin, J.}, \bibinfo{author}{Chang, M.},
  \bibinfo{author}{Lee, K.}, \& \bibinfo{author}{Toutanova, K.}
  (\bibinfo{year}{2019}).
\newblock \bibinfo{title}{{BERT:} pre-training of deep bidirectional
  transformers for language understanding}.
\newblock In \bibinfo{editor}{J.~Burstein}, \bibinfo{editor}{C.~Doran}, \&
  \bibinfo{editor}{T.~Solorio} (Eds.), {\it \bibinfo{booktitle}{Proceedings of
  the 2019 Conference of the North American Chapter of the Association for
  Computational Linguistics: Human Language Technologies, {NAACL-HLT} 2019,
  Minneapolis, MN, USA, June 2-7, 2019, Volume 1 (Long and Short Papers)}\/}
  (pp. \bibinfo{pages}{4171--4186}).
\newblock \bibinfo{publisher}{Association for Computational Linguistics}.
\newblock \DOIprefix\doi{10.18653/V1/N19-1423}.
\bibitem[{Feng et~al.(2020)Feng, Guo, Tang, Duan, Feng, Gong, Shou, Qin, Liu,
  Jiang \& Zhou}]{key-codebert}
\bibinfo{author}{Feng, Z.}, \bibinfo{author}{Guo, D.}, \bibinfo{author}{Tang,
  D.}, \bibinfo{author}{Duan, N.}, \bibinfo{author}{Feng, X.},
  \bibinfo{author}{Gong, M.}, \bibinfo{author}{Shou, L.}, \bibinfo{author}{Qin,
  B.}, \bibinfo{author}{Liu, T.}, \bibinfo{author}{Jiang, D.}, \&
  \bibinfo{author}{Zhou, M.} (\bibinfo{year}{2020}).
\newblock \bibinfo{title}{Codebert: {A} pre-trained model for programming and
  natural languages}.
\newblock In \bibinfo{editor}{T.~Cohn}, \bibinfo{editor}{Y.~He}, \&
  \bibinfo{editor}{Y.~Liu} (Eds.), {\it \bibinfo{booktitle}{Findings of the
  Association for Computational Linguistics: {EMNLP} 2020, Online Event, 16-20
  November 2020}\/} (pp. \bibinfo{pages}{1536--1547}).
\newblock \bibinfo{publisher}{Association for Computational Linguistics} volume
  \bibinfo{volume}{{EMNLP} 2020} of {\it \bibinfo{series}{Findings of
  {ACL}}\/}.
\newblock \DOIprefix\doi{10.18653/V1/2020.FINDINGS-EMNLP.139}.
\bibitem[{Fried et~al.(2023)Fried, Aghajanyan, Lin, Wang, Wallace, Shi, Zhong,
  Yih, Zettlemoyer \& Lewis}]{fried2022incoder}
\bibinfo{author}{Fried, D.}, \bibinfo{author}{Aghajanyan, A.},
  \bibinfo{author}{Lin, J.}, \bibinfo{author}{Wang, S.},
  \bibinfo{author}{Wallace, E.}, \bibinfo{author}{Shi, F.},
  \bibinfo{author}{Zhong, R.}, \bibinfo{author}{Yih, S.},
  \bibinfo{author}{Zettlemoyer, L.}, \& \bibinfo{author}{Lewis, M.}
  (\bibinfo{year}{2023}).
\newblock \bibinfo{title}{Incoder: {A} generative model for code infilling and
  synthesis}, .
\bibitem[{Guo et~al.(2022)Guo, Lu, Duan, Wang, Zhou \& Yin}]{guo2022unixcoder}
\bibinfo{author}{Guo, D.}, \bibinfo{author}{Lu, S.}, \bibinfo{author}{Duan,
  N.}, \bibinfo{author}{Wang, Y.}, \bibinfo{author}{Zhou, M.}, \&
  \bibinfo{author}{Yin, J.} (\bibinfo{year}{2022}).
\newblock \bibinfo{title}{Unixcoder: Unified cross-modal pre-training for code
  representation}, .
\newblock (pp. \bibinfo{pages}{7212--7225}).
  \DOIprefix\doi{10.18653/V1/2022.ACL-LONG.499}.
\bibitem[{Guo et~al.(2021)Guo, Ren, Lu, Feng, Tang, Liu, Zhou, Duan,
  Svyatkovskiy, Fu, Tufano, Deng, Clement, Drain, Sundaresan, Yin, Jiang \&
  Zhou}]{guo2020graphcodebert}
\bibinfo{author}{Guo, D.}, \bibinfo{author}{Ren, S.}, \bibinfo{author}{Lu, S.},
  \bibinfo{author}{Feng, Z.}, \bibinfo{author}{Tang, D.}, \bibinfo{author}{Liu,
  S.}, \bibinfo{author}{Zhou, L.}, \bibinfo{author}{Duan, N.},
  \bibinfo{author}{Svyatkovskiy, A.}, \bibinfo{author}{Fu, S.},
  \bibinfo{author}{Tufano, M.}, \bibinfo{author}{Deng, S.~K.},
  \bibinfo{author}{Clement, C.~B.}, \bibinfo{author}{Drain, D.},
  \bibinfo{author}{Sundaresan, N.}, \bibinfo{author}{Yin, J.},
  \bibinfo{author}{Jiang, D.}, \& \bibinfo{author}{Zhou, M.}
  (\bibinfo{year}{2021}).
\newblock \bibinfo{title}{Graphcodebert: Pre-training code representations with
  data flow}.
\newblock In {\it \bibinfo{booktitle}{9th International Conference on Learning
  Representations, {ICLR} 2021, Virtual Event, Austria, May 3-7, 2021}\/}.
\newblock \bibinfo{publisher}{OpenReview.net}.
\bibitem[{Haque et~al.(2022)Haque, Eberhart, Bansal \&
  McMillan}]{haque2022semantic}
\bibinfo{author}{Haque, S.}, \bibinfo{author}{Eberhart, Z.},
  \bibinfo{author}{Bansal, A.}, \& \bibinfo{author}{McMillan, C.}
  (\bibinfo{year}{2022}).
\newblock \bibinfo{title}{Semantic similarity metrics for evaluating source
  code summarization}.
\newblock In {\it \bibinfo{booktitle}{Proceedings of the 30th IEEE/ACM
  International Conference on Program Comprehension}\/} (pp.
  \bibinfo{pages}{36--47}).
\newblock \DOIprefix\doi{10.1145/3524610.3527909}.
\bibitem[{Hemberg et~al.(2024)Hemberg, Moskal \&
  O'Reilly}]{DBLP:journals/gpem/HembergMO24}
\bibinfo{author}{Hemberg, E.}, \bibinfo{author}{Moskal, S.}, \&
  \bibinfo{author}{O'Reilly, U.} (\bibinfo{year}{2024}).
\newblock \bibinfo{title}{Evolving code with a large language model}.
\newblock {\it \bibinfo{journal}{Genet. Program. Evolvable Mach.}\/},  {\it
  \bibinfo{volume}{25}\/}, \bibinfo{pages}{21}.
  \DOIprefix\doi{10.1007/S10710-024-09494-2}.
\bibitem[{Husein et~al.(2025)Husein, Aburajouh \&
  Catal}]{DBLP:journals/csi/HuseinAC25}
\bibinfo{author}{Husein, R.~A.}, \bibinfo{author}{Aburajouh, H.}, \&
  \bibinfo{author}{Catal, C.} (\bibinfo{year}{2025}).
\newblock \bibinfo{title}{Large language models for code completion: {A}
  systematic literature review}.
\newblock {\it \bibinfo{journal}{Comput. Stand. Interfaces}\/},  {\it
  \bibinfo{volume}{92}\/}, \bibinfo{pages}{103917}.
  \DOIprefix\doi{10.1016/J.CSI.2024.103917}.
\bibitem[{Kanade et~al.(2020)Kanade, Maniatis, Balakrishnan \&
  Shi}]{kanade2020learning}
\bibinfo{author}{Kanade, A.}, \bibinfo{author}{Maniatis, P.},
  \bibinfo{author}{Balakrishnan, G.}, \& \bibinfo{author}{Shi, K.}
  (\bibinfo{year}{2020}).
\newblock \bibinfo{title}{Learning and evaluating contextual embedding of
  source code}.
\newblock In {\it \bibinfo{booktitle}{Proceedings of the 37th International
  Conference on Machine Learning, {ICML} 2020, 13-18 July 2020, Virtual
  Event}\/} (pp. \bibinfo{pages}{5110--5121}).
\newblock \bibinfo{publisher}{{PMLR}} volume \bibinfo{volume}{119} of {\it
  \bibinfo{series}{Proceedings of Machine Learning Research}\/}.
\newblock \URLprefix \url{http://proceedings.mlr.press/v119/kanade20a.html}.
\bibitem[{Karmakar \& Robbes(2021)}]{karmakar2021pre}
\bibinfo{author}{Karmakar, A.}, \& \bibinfo{author}{Robbes, R.}
  (\bibinfo{year}{2021}).
\newblock \bibinfo{title}{What do pre-trained code models know about code?}
\newblock In {\it \bibinfo{booktitle}{2021 36th IEEE/ACM International
  Conference on Automated Software Engineering (ASE)}\/} (pp.
  \bibinfo{pages}{1332--1336}).
\newblock \bibinfo{organization}{IEEE}.
\newblock \DOIprefix\doi{10.1109/ASE51524.2021.9678927}.
\bibitem[{Karnalim et~al.(2019)Karnalim, Budi, Toba \&
  Joy}]{karnalim2019source}
\bibinfo{author}{Karnalim, O.}, \bibinfo{author}{Budi, S.},
  \bibinfo{author}{Toba, H.}, \& \bibinfo{author}{Joy, M.}
  (\bibinfo{year}{2019}).
\newblock \bibinfo{title}{Source code plagiarism detection in academia with
  information retrieval: Dataset and the observation}.
\newblock {\it \bibinfo{journal}{Informatics in Education}\/},  {\it
  \bibinfo{volume}{18}\/}, \bibinfo{pages}{321--344}.
  \DOIprefix\doi{10.15388/INFEDU.2019.15}.
\bibitem[{Kaur \& Rattan(2023)}]{key-KaurR23}
\bibinfo{author}{Kaur, M.}, \& \bibinfo{author}{Rattan, D.}
  (\bibinfo{year}{2023}).
\newblock \bibinfo{title}{A systematic literature review on the use of machine
  learning in code clone research}.
\newblock {\it \bibinfo{journal}{Comput. Sci. Rev.}\/},  {\it
  \bibinfo{volume}{47}\/}, \bibinfo{pages}{100528}.
  \DOIprefix\doi{10.1016/J.COSREV.2022.100528}.
\bibitem[{Li et~al.(2023)Li, Allal, Zi, Muennighoff, Kocetkov, Mou, Marone,
  Akiki, Li, Chim et~al.}]{li2023starcoder}
\bibinfo{author}{Li, R.}, \bibinfo{author}{Allal, L.~B.}, \bibinfo{author}{Zi,
  Y.}, \bibinfo{author}{Muennighoff, N.}, \bibinfo{author}{Kocetkov, D.},
  \bibinfo{author}{Mou, C.}, \bibinfo{author}{Marone, M.},
  \bibinfo{author}{Akiki, C.}, \bibinfo{author}{Li, J.}, \bibinfo{author}{Chim,
  J.} et~al. (\bibinfo{year}{2023}).
\newblock \bibinfo{title}{Starcoder: may the source be with you!}
\newblock {\it \bibinfo{journal}{Trans. Mach. Learn. Res.}\/},  {\it
  \bibinfo{volume}{2023}\/}.
\bibitem[{Lozhkov et~al.(2024)Lozhkov, Li, Allal, Cassano, Lamy-Poirier, Tazi,
  Tang, Pykhtar, Liu, Wei et~al.}]{lozhkov2024starcoder}
\bibinfo{author}{Lozhkov, A.}, \bibinfo{author}{Li, R.},
  \bibinfo{author}{Allal, L.~B.}, \bibinfo{author}{Cassano, F.},
  \bibinfo{author}{Lamy-Poirier, J.}, \bibinfo{author}{Tazi, N.},
  \bibinfo{author}{Tang, A.}, \bibinfo{author}{Pykhtar, D.},
  \bibinfo{author}{Liu, J.}, \bibinfo{author}{Wei, Y.} et~al.
  (\bibinfo{year}{2024}).
\newblock \bibinfo{title}{Starcoder 2 and the stack v2: The next generation}.
\newblock {\it \bibinfo{journal}{arXiv preprint arXiv:2402.19173}\/}, .
\bibitem[{Martinez-Gil(2023)}]{key-martinez-ijseke}
\bibinfo{author}{Martinez-Gil, J.} (\bibinfo{year}{2023}).
\newblock \bibinfo{title}{A comparative study of ensemble techniques based on
  genetic programming: {A} case study in semantic similarity assessment}.
\newblock {\it \bibinfo{journal}{Int. J. Softw. Eng. Knowl. Eng.}\/},  {\it
  \bibinfo{volume}{33}\/}, \bibinfo{pages}{289--312}.
  \DOIprefix\doi{10.1142/S0218194022500772}.
\bibitem[{Martinez-Gil(2024{\natexlab{a}})}]{key-martinez-codesim}
\bibinfo{author}{Martinez-Gil, J.} (\bibinfo{year}{2024}{\natexlab{a}}).
\newblock \bibinfo{title}{Advanced detection of source code clones via an
  ensemble of unsupervised similarity measures}.
\newblock {\it \bibinfo{journal}{arXiv preprint arXiv:2405.02095}\/}, .
  \DOIprefix\doi{10.48550/arXiv.2405.02095}.
\bibitem[{Martinez-Gil(2024{\natexlab{b}})}]{martinez2024augmenting}
\bibinfo{author}{Martinez-Gil, J.} (\bibinfo{year}{2024}{\natexlab{b}}).
\newblock \bibinfo{title}{Augmenting the interpretability of graphcodebert for
  code similarity tasks}.
\newblock {\it \bibinfo{journal}{arXiv preprint arXiv:2410.05275}\/}, .
\bibitem[{Martinez-Gil(2024{\natexlab{c}})}]{key-martinez-swqd}
\bibinfo{author}{Martinez-Gil, J.} (\bibinfo{year}{2024}{\natexlab{c}}).
\newblock \bibinfo{title}{Source code clone detection using unsupervised
  similarity measures}.
\newblock In \bibinfo{editor}{P.~Bludau}, \bibinfo{editor}{R.~Ramler},
  \bibinfo{editor}{D.~Winkler}, \& \bibinfo{editor}{J.~Bergsmann} (Eds.), {\it
  \bibinfo{booktitle}{Software Quality as a Foundation for Security - 16th
  International Conference on Software Quality, {SWQD} 2024, Vienna, Austria,
  April 23-25, 2024, Proceedings}\/} (pp. \bibinfo{pages}{21--37}).
\newblock \bibinfo{publisher}{Springer} volume \bibinfo{volume}{505} of {\it
  \bibinfo{series}{Lecture Notes in Business Information Processing}\/}.
\newblock \DOIprefix\doi{10.1007/978-3-031-56281-5\_2}.
\bibitem[{Martinez-Gil \& Yin(2024)}]{martinez2024evaluation}
\bibinfo{author}{Martinez-Gil, J.}, \& \bibinfo{author}{Yin, S.}
  (\bibinfo{year}{2024}).
\newblock \bibinfo{title}{Evaluation of code similarity search strategies in
  large-scale codebases}.
\newblock In {\it \bibinfo{booktitle}{Transactions on Large-Scale Data-and
  Knowledge-Centered Systems LVII}\/} (pp. \bibinfo{pages}{99--113}).
\newblock \bibinfo{publisher}{Springer}.
\newblock \DOIprefix\doi{10.1007/978-3-662-70140-9\_4}.
\bibitem[{Mou et~al.(2016)Mou, Li, Zhang, Wang \& Jin}]{mou2016convolutional}
\bibinfo{author}{Mou, L.}, \bibinfo{author}{Li, G.}, \bibinfo{author}{Zhang,
  L.}, \bibinfo{author}{Wang, T.}, \& \bibinfo{author}{Jin, Z.}
  (\bibinfo{year}{2016}).
\newblock \bibinfo{title}{Convolutional neural networks over tree structures
  for programming language processing}.
\newblock In \bibinfo{editor}{D.~Schuurmans}, \& \bibinfo{editor}{M.~P.
  Wellman} (Eds.), {\it \bibinfo{booktitle}{Proceedings of the Thirtieth {AAAI}
  Conference on Artificial Intelligence, February 12-17, 2016, Phoenix,
  Arizona, {USA}}\/} (pp. \bibinfo{pages}{1287--1293}).
\newblock \bibinfo{publisher}{{AAAI} Press}.
\newblock \DOIprefix\doi{10.1609/AAAI.V30I1.10139}.
\bibitem[{Nasrabadi et~al.(2023)Nasrabadi, Parsa, Ramezani, Roy \&
  Ekhtiarzadeh}]{zakeri2023systematic}
\bibinfo{author}{Nasrabadi, M.~Z.}, \bibinfo{author}{Parsa, S.},
  \bibinfo{author}{Ramezani, M.}, \bibinfo{author}{Roy, C.}, \&
  \bibinfo{author}{Ekhtiarzadeh, M.} (\bibinfo{year}{2023}).
\newblock \bibinfo{title}{A systematic literature review on source code
  similarity measurement and clone detection: Techniques, applications, and
  challenges}.
\newblock {\it \bibinfo{journal}{J. Syst. Softw.}\/},  {\it
  \bibinfo{volume}{204}\/}, \bibinfo{pages}{111796}.
  \DOIprefix\doi{10.1016/J.JSS.2023.111796}.
\bibitem[{Pham et~al.(2024)Pham, Ha, Tong, Hoang, Tran \&
  Le}]{DBLP:journals/access/PhamHTHTL24}
\bibinfo{author}{Pham, H.}, \bibinfo{author}{Ha, H.}, \bibinfo{author}{Tong,
  V.}, \bibinfo{author}{Hoang, D.}, \bibinfo{author}{Tran, D.}, \&
  \bibinfo{author}{Le, T.~N.} (\bibinfo{year}{2024}).
\newblock \bibinfo{title}{{MAGECODE:} machine-generated code detection method
  using large language models}.
\newblock {\it \bibinfo{journal}{{IEEE} Access}\/},  {\it
  \bibinfo{volume}{12}\/}, \bibinfo{pages}{190186--190202}.
  \DOIprefix\doi{10.1109/ACCESS.2024.3509987}.
\bibitem[{PoolC(2022)}]{poolc}
\bibinfo{author}{PoolC} (\bibinfo{year}{2022}).
\newblock \bibinfo{title}{1-fold-clone-detection-600k-5fold}.
\newblock
  \bibinfo{howpublished}{\url{https://huggingface.co/datasets/PoolC/1-fold-clone-detection-600k-5fold}}.
\bibitem[{Ramler et~al.(2024)Ramler, Moser, Fischer, Nissl \&
  Heinzl}]{DBLP:conf/llm4code/RamlerM0NH24}
\bibinfo{author}{Ramler, R.}, \bibinfo{author}{Moser, M.},
  \bibinfo{author}{Fischer, L.}, \bibinfo{author}{Nissl, M.}, \&
  \bibinfo{author}{Heinzl, R.} (\bibinfo{year}{2024}).
\newblock \bibinfo{title}{Industrial experience report on ai-assisted coding in
  professional software development}.
\newblock In {\it \bibinfo{booktitle}{LLM4CODE\@ICSE}\/} (pp.
  \bibinfo{pages}{1--7}).
\newblock \DOIprefix\doi{10.1145/3643795.3648377}.
\bibitem[{Rattan et~al.(2013)Rattan, Bhatia \& Singh}]{rattan2013software}
\bibinfo{author}{Rattan, D.}, \bibinfo{author}{Bhatia, R.}, \&
  \bibinfo{author}{Singh, M.} (\bibinfo{year}{2013}).
\newblock \bibinfo{title}{Software clone detection: A systematic review}.
\newblock {\it \bibinfo{journal}{Information and Software Technology}\/},  {\it
  \bibinfo{volume}{55}\/}, \bibinfo{pages}{1165--1199}.
  \DOIprefix\doi{10.1016/J.INFSOF.2013.01.008}.
\bibitem[{Roziere et~al.(2023)Roziere, Gehring, Gloeckle, Sootla, Gat, Tan,
  Adi, Liu, Sauvestre, Remez et~al.}]{roziere2023code}
\bibinfo{author}{Roziere, B.}, \bibinfo{author}{Gehring, J.},
  \bibinfo{author}{Gloeckle, F.}, \bibinfo{author}{Sootla, S.},
  \bibinfo{author}{Gat, I.}, \bibinfo{author}{Tan, X.~E.},
  \bibinfo{author}{Adi, Y.}, \bibinfo{author}{Liu, J.},
  \bibinfo{author}{Sauvestre, R.}, \bibinfo{author}{Remez, T.} et~al.
  (\bibinfo{year}{2023}).
\newblock \bibinfo{title}{Code llama: Open foundation models for code}.
\newblock {\it \bibinfo{journal}{arXiv preprint arXiv:2308.12950}\/}, .
\bibitem[{Svajlenko \& Roy(2021)}]{svajlenko2021bigclonebench}
\bibinfo{author}{Svajlenko, J.}, \& \bibinfo{author}{Roy, C.~K.}
  (\bibinfo{year}{2021}).
\newblock \bibinfo{title}{Bigclonebench}.
\newblock {\it \bibinfo{journal}{Code Clone Analysis: Research, Tools, and
  Practices}\/},  (pp. \bibinfo{pages}{93--105}).
  \DOIprefix\doi{10.1007/978-981-16-1927-4\_7}.
\bibitem[{Tunstall et~al.(2022)Tunstall, Von~Werra \&
  Wolf}]{tunstall2022natural}
\bibinfo{author}{Tunstall, L.}, \bibinfo{author}{Von~Werra, L.}, \&
  \bibinfo{author}{Wolf, T.} (\bibinfo{year}{2022}).
\newblock {\it \bibinfo{title}{Natural language processing with
  transformers}\/}.
\newblock \bibinfo{publisher}{O'Reilly Media, Inc.}
\bibitem[{Wang et~al.(2021)Wang, Wang, Joty \& Hoi}]{wang2021codet5}
\bibinfo{author}{Wang, Y.}, \bibinfo{author}{Wang, W.}, \bibinfo{author}{Joty,
  S.~R.}, \& \bibinfo{author}{Hoi, S. C.~H.} (\bibinfo{year}{2021}).
\newblock \bibinfo{title}{Codet5: Identifier-aware unified pre-trained
  encoder-decoder models for code understanding and generation}, .
\newblock (pp. \bibinfo{pages}{8696--8708}).
  \DOIprefix\doi{10.18653/V1/2021.EMNLP-MAIN.685}.
\bibitem[{Xu et~al.(2022)Xu, Alon, Neubig \& Hellendoorn}]{xu2022systematic}
\bibinfo{author}{Xu, F.~F.}, \bibinfo{author}{Alon, U.},
  \bibinfo{author}{Neubig, G.}, \& \bibinfo{author}{Hellendoorn, V.~J.}
  (\bibinfo{year}{2022}).
\newblock \bibinfo{title}{A systematic evaluation of large language models of
  code}, .
\newblock (pp. \bibinfo{pages}{1--10}).
  \DOIprefix\doi{10.1145/3520312.3534862}.
\bibitem[{Zhang et~al.(2024)Zhang, Chen, Cao, Chen \&
  Zhou}]{DBLP:journals/ijseke/ZhangCCCZ24}
\bibinfo{author}{Zhang, X.}, \bibinfo{author}{Chen, Z.}, \bibinfo{author}{Cao,
  Y.}, \bibinfo{author}{Chen, L.}, \& \bibinfo{author}{Zhou, Y.}
  (\bibinfo{year}{2024}).
\newblock \bibinfo{title}{Multi-intent inline code comment generation via large
  language model}.
\newblock {\it \bibinfo{journal}{Int. J. Softw. Eng. Knowl. Eng.}\/},  {\it
  \bibinfo{volume}{34}\/}, \bibinfo{pages}{845--868}.
  \DOIprefix\doi{10.1142/S0218194024500050}.
\bibitem[{Zhang et~al.(2023)Zhang, Li, Zhang, Long, Xie, Zhang \&
  Zhang}]{zhang2023language}
\bibinfo{author}{Zhang, X.}, \bibinfo{author}{Li, Z.}, \bibinfo{author}{Zhang,
  Y.}, \bibinfo{author}{Long, D.}, \bibinfo{author}{Xie, P.},
  \bibinfo{author}{Zhang, M.}, \& \bibinfo{author}{Zhang, M.}
  (\bibinfo{year}{2023}).
\newblock \bibinfo{title}{Language models are universal embedders}.
\newblock {\it \bibinfo{journal}{arXiv preprint arXiv:2310.08232}\/}, .
\bibitem[{Zhang \& Saber(2025)}]{zhang2025machine}
\bibinfo{author}{Zhang, Z.}, \& \bibinfo{author}{Saber, T.}
  (\bibinfo{year}{2025}).
\newblock \bibinfo{title}{Machine learning approaches to code similarity
  measurement: A systematic review}.
\newblock {\it \bibinfo{journal}{IEEE Access}\/}, .
  \DOIprefix\doi{10.1109/ACCESS.2025.3553392}.
\bibitem[{Zhao \& Huang(2018)}]{zhao2018deepsim}
\bibinfo{author}{Zhao, G.}, \& \bibinfo{author}{Huang, J.}
  (\bibinfo{year}{2018}).
\newblock \bibinfo{title}{Deepsim: deep learning code functional similarity}.
\newblock In \bibinfo{editor}{G.~T. Leavens}, \bibinfo{editor}{A.~Garcia}, \&
  \bibinfo{editor}{C.~S. Pasareanu} (Eds.), {\it
  \bibinfo{booktitle}{Proceedings of the 2018 {ACM} Joint Meeting on European
  Software Engineering Conference and Symposium on the Foundations of Software
  Engineering, {ESEC/SIGSOFT} {FSE} 2018, Lake Buena Vista, FL, USA, November
  04-09, 2018}\/} (pp. \bibinfo{pages}{141--151}).
\newblock \bibinfo{publisher}{{ACM}}.
\newblock \DOIprefix\doi{10.1145/3236024.3236068}.
\bibitem[{Zhao et~al.(2025)Zhao, Luo, Tian, Lin, Yan, Li \& Ma}]{ZhaoLT0YL025}
\bibinfo{author}{Zhao, Y.}, \bibinfo{author}{Luo, Z.}, \bibinfo{author}{Tian,
  Y.}, \bibinfo{author}{Lin, H.}, \bibinfo{author}{Yan, W.},
  \bibinfo{author}{Li, A.}, \& \bibinfo{author}{Ma, J.} (\bibinfo{year}{2025}).
\newblock \bibinfo{title}{Codejudge-eval: Can large language models be good
  judges in code understanding?}
\newblock In \bibinfo{editor}{O.~Rambow}, \bibinfo{editor}{L.~Wanner},
  \bibinfo{editor}{M.~Apidianaki}, \bibinfo{editor}{H.~Al{-}Khalifa},
  \bibinfo{editor}{B.~D. Eugenio}, \& \bibinfo{editor}{S.~Schockaert} (Eds.),
  {\it \bibinfo{booktitle}{Proceedings of the 31st International Conference on
  Computational Linguistics, {COLING} 2025, Abu Dhabi, UAE, January 19-24,
  2025}\/} (pp. \bibinfo{pages}{73--95}).
\newblock \bibinfo{publisher}{Association for Computational Linguistics}.
\newblock \URLprefix \url{https://aclanthology.org/2025.coling-main.7/}.

\end{thebibliography}

\end{document}